# CONCEPTUALIZING A MULTI-SIDED PLATFORM FOR CLOUD COMPUTING RESOURCE TRADING

*Research Paper*

Franziska, Haller, Karlsruhe Institute of Technology & IBM, Germany, franziska.haller@alumni.kit.edu

Max, Schemmer, Karlsruhe Institute of Technology, Germany, max.schemmer@kit.edu

Niklas, Kühl, University of Bayreuth, Germany, kuehl@uni-bayreuth.de

Carsten, Holtmann, IBM, Germany, carsten.holtmann@de.ibm.com

## Abstract

*Cost-effective and responsible use of cloud computing resources (CCR) is on the business agenda of many companies. Despite this strategic goal, two geopolitical strategy decisions mainly influence the continuous existence of overcapacity: Europe's General Data Protection Regulation and the US's Cloud Act. Given the circumstances, a typical data center produces approximately 30% overcapacity annually. This overcapacity has severe environmental and economic consequences. Our work addresses this overcapacity by proposing a multi-sided platform for CCR trading. We initiate our research by conducting a literature review to explore the existing body of knowledge which indicates a lack of recent and evaluated platform design knowledge for CCR trading. We address this research gap by deriving design requirements and design principles. We instantiate and evaluate the design knowledge in a respective platform framework. Thus, we contribute to research and practice by deriving and evaluating design knowledge and proposing a platform framework.*

## 1    Introduction

Cost-effective and responsible use of cloud computing resources (CCR) is a goal for companies of all sizes (Comella-Dorda *et al.*, 2018; Varghese and Buyya, 2017). Despite this strategic goal, two geopolitical strategy decisions have mainly influenced the continuous existence of overcapacity: Europe's General Data Protection Regulation (GDPR) and the US's Cloud Act. Given the circumstances, a typical data center produces an estimated 30% overcapacity annually (Ngoko and Cerin, 2017). This overcapacity has severe consequences from multiple perspectives.

From an environmental perspective, this overcapacity needs to be addressed as servers utilized at 10% capacity or lower, still use up to 60% of their maximum power even while performing little to no calculations (Whitney and Delforge, 2014). Trends show, that energy consumption will increase steadily until 2030, with hyperscale data centers having doubled their energy demand between 2015-2019 (European Commission, 2020; Statista, 2023). From an economic perspective, market dynamics indicate a continuous growth in computing demand (Liu *et al.*, 2020). Simultaneously, the supply side is stressed due to shortages of necessary resources such as semiconductors, requiring companies to reconsider their cloud sourcing decisions (Schneider and Sunyaev, 2016). With a shortage in semiconductors, the supply of additional CCR decreases. Thus, the need for effectively using overcapacity in CCR is prevalent already. Furthermore, idle CCR produce significant opportunity costs (Statista, 2022; Whitney and Delforge, 2014). Selling overcapacity would allow companies to decrease these costs—a concept that has been implemented in a similar way in the energy industry for non-storable energy, e.g., wind power (Dinther *et al.*, 2021). Since there are hundreds of millions of unit server racks, this has massive energy and economic potential (Maguranis *et al.*, 2021).

Research has mainly discussed three solution approaches to address overcapacity 1) optimization of CCR planning (Mann, 2015), 2) purchase of additional CCR to cover peak utilization with third-party resources (Arora *et al.*, 2020), also known as cloud sourcing (Muhic and Johansson, 2014), and 3) sale of CCR overcapacity. Literature addresses the optimization of CCR planning extensively (Mann, 2015). Companies' sourcing of CCR for peak load events is a standard business solution (Arora, 2020; Muhic and Johansson, 2014). However, practice has shown that 95% of small, medium, corporate, and multi-tenant operations still produce significant overcapacity despite the possibility of buying additional third-party resources for peak load events (Whitney and Delforge, 2014). The reasons for that are multifold, e.g., data security concerns (Kerkmann and Müller, 2020). Hence, in this article, we address the third scenario, the sale of CCR overcapacity via a multi-sided platform (MSP), since research on current and evaluated platform design knowledge for CCR trading platforms is exceptionally scarce (Alrawahi and Lee, 2012). In doing so, we expand the classic cloud sourcing process by a CCR trading process.

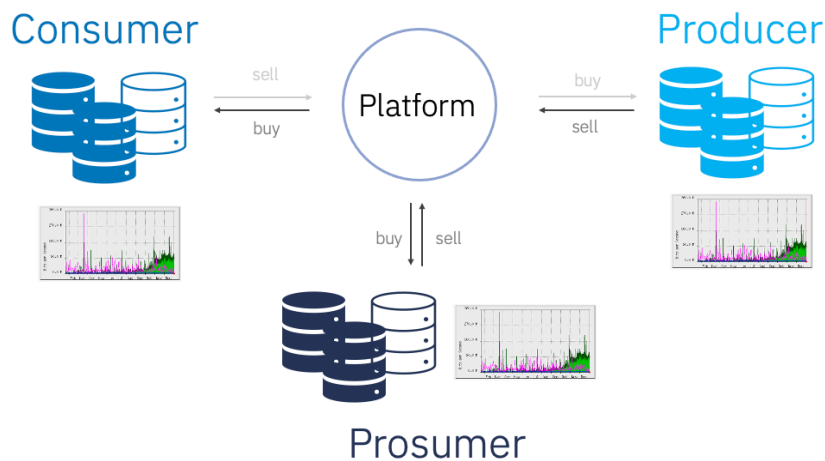

*Figure 1.*      *Multi-sided platform for cloud computing resource trading (based on Hagiu and Wright, 2015)*

MSPs connect two or more interdependent user groups by playing an intermediation role (Evans and Schmalensee, 2019). Based on the foundations of MSP research, we develop a platform framework to exchange overcapacity between providers, consumers, and prosumers—entities providing and sourcing overcapacity. For the remainder of this article, the traded overcapacity refers to idle, obsolete, or unused CCR. Hereby, CCR refers to central processing unit (CPU) capacity—virtualized shares of physical CPUs.

Our platform approach addresses a lack of recent and evaluated design knowledge for MSPs for CCR trading. Furthermore, it provides a solution to promote innovation and competitiveness in the cloud market—a goal regulators worldwide pursue vigorously with the support of companies of all industries (Draheim, 2021; Schwartz, 2018). Hence, this work addresses the need for research and practice by providing evaluated design knowledge.

To do so, this article addresses the following research questions:

**RQ 1:** *What requirements can be derived for a multi-sided platform for cloud computing trading?*

**RQ 2:** *What principles can guide the development of a multi-sided platform for cloud computing trading?*

To answer these research questions, we conduct a Design Science Research (DSR) project and derive design knowledge for CCR trading (Hevner and Chatterjee, 2010). Based on DSR, we derive design requirements (DR) and design principles (DP) to develop a generalized framework for a CCR trading platform. After that, we apply the framework to the European cloud market. Previous trading platforms faced the challenge of lacking platform governance and standards for CCR (Altmann et al., 2008; Khurana et al., 2010). Current European initiatives —namely Gaia-X—might have the potential to resolve that issue. Gaia-X is a European initiative developing standards and requirements in line with EU regulations for data and infrastructure (Braud et al., 2021). Based on Gaia-X, we discuss an instantiated version of the framework that we call CCP-X (Cloud Computing Platform enabled by Gaia-X). In doing so, we address Gaia-X's call for participation in strengthening the competitiveness of the European cloud market (Braud et al., 2021). Further instantiations of the framework could be addressed in future work.

Our contribution to research and practice is threefold: First, we derive DR through a structured literature review (SLR) and an interview study with domain experts. Second, we deduct DP based on conceptual foundations from energy markets. The energy market serves as an analogy since it integrated distributed energy resources via peer-to-peer (P2P) energy trading platforms into its traditional central market structure (Renewable Energy Agency, 2020). Third, we develop, instantiate and evalaluate a trading platform framework for distributed CCR.

## 2    Methodology

Our research follows the Design Science Research approach (Hevner *et al.*, 2004; March and Smith, 1995). In particular, we follow the three-cycle Design Science Research guidelines from Hevner (2007)—the relevance, rigor, and design cycle in the style of (Laing and Kühl, 2020).

The **rigor cycle** provides the research with grounding theories and methodology from the knowledge base (input from the theoretical knowledge base). Our rigor cycle has two purposes. First, it ensures the theoretical research gap. Second, we analyze literature in a related domain dealing with non-storable resources to ground our design. To provide a rigourous investigation of the research gap, we followed the methodology outlined by Vom Brocke et al. (2009) and conduct an SLR. In line with the outlined methodology, we develop a search string to ensure a comprehensible literature research: *TITLE-ABS-KEY (("cloud exchange" OR "cloud marketplace" OR "cloud market" OR "cloud platform") AND "cloud infrastructure")*. Literature on MSPs and cloud platforms spans different journals, publishers, and conferences. Hence, we decided on the SCOPUS database, a large single abstract and indexing database, which provides linked scholarly literature across a variety of disciplines (Burnham, 2006). Based on the search string, we performed a rigorous keyword-based search. We initially identified 328 articles through the keyword-based search. Screening the articles, we identify 61 potentially eligible full-text articles. Following, qualitative analysis, including a forward and backward search, results in 24 eligible articles. We analyzed these articles in-depth realizing, that the body of knowledge in this domain is very scarce.

The **relevance cycle** provides the research with environmental requirements and ensures field testing (input from the practical knowledge base). In the relevance cycle, problems of an actual application environment should be identified and defined (Hevner, 2007). To do so, we decided on a semi-structured explorative interview study since interviews allow us to collect first-hand opinions and impressions about the application domain and generate theory by collecting insightful data from experts with different roles and expertise (Martin *et al.*, 2012; Myers and Newman, 2007). This allows us to address weaknesses of prior research. For the semi-structured interviews, we followed the guidelines by Whiting (2008). In total, we conducted interviews with 13 experts, including 4 C-level executives. We invited experts in the areas of technology, industry, and regulation. All interviews were recorded and transcribed. The interviews lasted between 25 minutes and 60 minutes. To analyze the interview transcripts and documentation, we followed an inductive open coding approach outlined by Mayring (2010). Our analysis resulted in a set of 297 first-order codes. We re-evaluated the first-order codes to generate exhaustive and mutually exclusive categories. Re-evaluation allowed us to consolidate several first-order codes, resulting in 51 second-order codes (Cassell and Symon, 2014). We then grouped the second-order codes into seven categories, with each category representing one design requirement. Based on the second-order codes, the inter-coder reliability of the two main researchers measured by Cohen's kappa was initially 40.6 percent. After discussing the codes the inter-coder reliability improved to 88.9 percent. The acceptance criteria in the form of domain requirements for the evaluation are derived by answering the question whether the artifact improved the environment (Hevner, 2007; Landwehr *et al.*, 2022).

| ID | Position | Stakeholder group | Company size |
|---|---|---|---|
| Alpha | CTO | Technology | >340.000 (Company A) |
| Beta | Gaia-X Expert | Regulation | >70 (Company B) |
| Gamma | Cloud Commodity Expert | Regulation | >50 (Company C) |
| Delta | C-Level Cloud Exchange Expert | Industry | >15 (Company D) |
| Epsilon | C-Level Cloud Exchange Expert | Industry | >15 (Company D) |
| Zêta | Cloud Expert | Technology | >340.000 (Company A) |
| Êta | Industry Expert | Industry | >340.000 (Company A) |
| Thêta | C-Level Data Center Expert | Industry | >10 (Company E) |
| Iota | Cloud Expert | Technology | >340.000 (Company A) |
| Kappa | Cloud Exchange Expert | Technology | >340.000 (Company A) |
| Lambda | Infrastructure Expert | Technology | >340.000 (Company A) |
| Mu | Cloud Expert | Regulation | >340.000 (Company A) |
| Nu | Gaia-X Expert | Regulation | >70 (Company B) |

*Table 1.        Explorative interview study sampling*

A **design cycle** (DC) incorporates the design, development, and evaluation of artifacts. Our DC follows the steps awareness of problem, suggestion, development, demonstration, and evaluation, based on Kuechler and Vaishnavi (2008) and Meth et al. (2015). In the **awareness of problem** step, we ensure the relevance of the problem we want to address in an environmental, macroeconomic, microeconomic, and geopolitical context. In the **suggestion** phase, we derive six DP based on the theoretical knowledge base of the energy market. Subsequently, in the **development** phase, we conduct an expert workshop based on the methodology outlined by Thoring, Mueller and Badke-Schaub (2020) with four highly experienced participants from research and industry. In the workshop, we map the abstract DP into a generalized framework for CCR trading platforms based on the research of Hagiu and Wright (2015). In the **demonstration** phase, we instantiate the framework "CCP-X" by applying it to the European cloud market and the Gaia-X infrastructure based on Gaia-X publications. Lastly, in our summative ex-post **evaluation** we evaluate our instantiated framework CCP-X within a focus group of four domain experts from the Gaia-X steering committee. We select our participants based on theoretical sampling by Coyne (1997) to guarantee a certain level of experience needed to foster an open discussion (Tremblay *et al.*, 2010). Thus, we provide a suitable context to analyze the DPs' effectiveness in addressing the identified DR and the general feasibility of the framework. Furthermore, we evaluate the feasibility of the framework as a Gaia-X use case. This approach suits our research for several reasons. First, this approach allows us to interact directly with experts familiar with the domain and clarify potential questions regarding the DP. Second, the flexibility of the method allows us to discuss additional topics with the experts, such as the instantiation of the platform within Gaia-X's consortiums. Last, the format enables us to present the platform and DP visually.

Figure 2 visualizes the interplay of the three cycles, including the objective, the method, and our result.

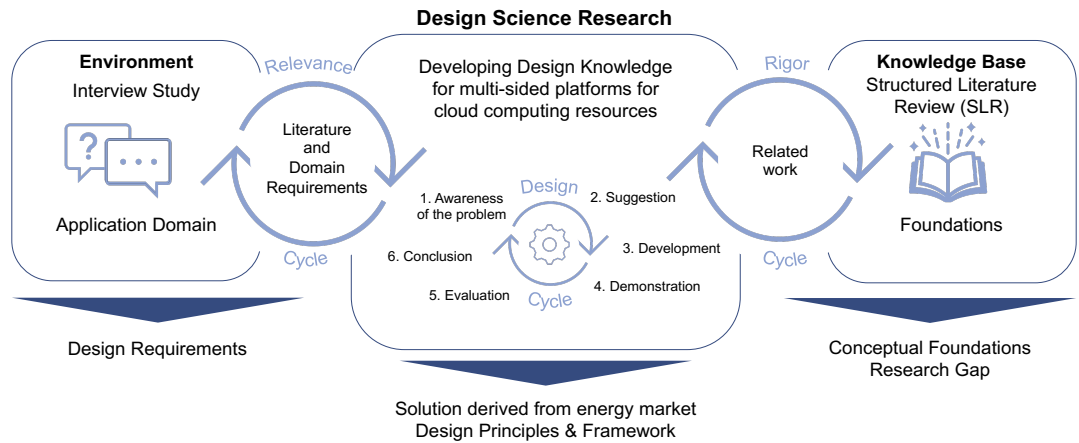

*Figure 2.*      *Design science research approach following Hevner and Chatterjee (2010).*

# 3    Rigor Cycle: Foundations & Related Work

In the following, we provide an overview of related work with regard to CCR trading platforms and introduce energy trading platforms which we use as inspirational input for our design cycle.

**CCR trading platforms.** Definitions of MSPs are diverse and have different areas of focus. Within this article, we define MSPs as outlined by Hagiu and Wright (2015), who identify two elements that are more critical than any other requirement: 1) *MSPs "enable direct interaction between two or more distinct sides"* (Hagiu and Wright, 2015, p. 163) and 2) *"Each side is affiliated with the platform"* (Hagiu & Wright, 2015, p.163). Enabling direct interaction allows two or more distinct sides to be in charge of the conditions of the interaction instead of the platform deciding on the conditions (Hagiu & Wright, 2015). In the case of a CCR trading platform, conditions might refer to the pricing or delivery of the goods.

The idea of sharing CCR overcapacity has created several market concepts. In the past, research has focused on resource allocation in the grid, also known as grid computing (Schnizler, 2007). However, among others, a lack of platform governance and missing standardization of CCR prevented the approach from establishing itself in the market (Altmann *et al.*, 2008; Khurana *et al.*, 2010). More recent platform approaches are *Crowdcloud* (Hosseini *et al.*, 2019), *Intercloud* (Buyya *et al.*, 2010), *OCX* ((Azer Bestavros, 2014), *Mandi* (Garg *et al.*, 2013), *Deutsche Börse Cloud Exchange (DBCE)* (Hagemeier, 2014), and a multi-attribute combinatorial marketplace for cloud resources (Alrawahi and Lee, 2012). *Crowdcloud*, *Intercloud*, and *OCX* highlight the importance of many stakeholders participating in implementing and operating the cloud rather than just a single provider. These platforms allow participants to take on the role of providers and consumers, thus opening the cloud market to a broader range of participants, including small and medium-sized businesses (Garg et al., 2013; Hosseini, 2019). *Mandi* provides a platform allowing the simultaneous coexistence of multiple trading negotiations (Garg et al., 2013), while *DBCE*, initiated in 2015, aimed at commoditizing cloud infrastructure as a service (IaaS) to make it tradeable on the *DBCE* platform (Hagemeier, 2014). Lastly, the multi-attribute combinatorial marketplace differentiates itself by proposing a marketplace focusing on the heterogeneity of cloud resources rather than unifying the provided resources into one resource (Alrawahi and Lee, 2012).

Our proposed platform framework takes existing research one step further by conducting the following steps: First, we base the platform approach on an interview study taking current market needs and market trends into consideration. Second, based on the generated data, explicit design knowledge for a CCR trading platform is derived and evaluated.

**Energy trading platforms.** As our SLR shows, little recent and evaluated research exists on MSPs for CCR trading. Therefore, we broaden our view and focus on a similar kind of platform and its surrounding market structure—P2P energy trading platforms. The cloud market shares many characteristics with the traditional energy market. The cloud market is centralized, with few providers and many consumers (Salehi & Makiyan, 2021). A CCR trading platform would allow new cloud providers to enter the market, offering overcapacity at a rate consumers are willing to pay. The trading platform is linked to traditional cloud providers, so-called hyperscalers, to provide reliable CCR availability. Hyperscalers would take on the role of a "residual balancer" since overcapacity is unpredictable, intermittent, and non-storable due to changing peak events.

Traditionally, the energy market had few providers and many consumers, resulting in a central structure. Few large provider plants generate electricity, transported via transmission and distribution networks to the customer. This unidirectional structure positions the consumer at the end of the supply chain purchasing electricity from utilities or retailers through fixed tariffs (Renewable Energy Agency, 2020).

With the emergence of decentralized energy resources, new active participants entered the electricity market, such as rooftop solar PV installations, micro wind turbines, and battery energy storage systems. The occurrence of those distributed energy resources forced the traditional electricity market to adapt its market structure (Renewable Energy Agency, 2020). New market players, such as prosumers, aggregators, and active consumers, needed to be incorporated into the existing electricity market structure. Prosumers are electricity consumers who were passive in the past and can actively contribute to the electricity market now by consuming and producing electricity (Parag and Sovacool, 2016).

Different P2P electricity trading platforms have incorporated distributed energy resources and new market participants into the traditional electricity market (Renewable Energy Agency, 2020). P2P electricity trading is a business model where consumers, producers, and prosumers meet through an interconnected MSP, allowing platform participants to meet directly, without an intermediary (Park and Yong, 2017). The platform takes on the role of an online marketplace where electricity producers sell their electricity at an individual price to consumers willing to pay that price. The platform is connected to the primary electricity grid to provide reliable electricity provision. The electricity grid takes on the role of a "residual balancer" since distributed energy resources are unpredictable, non-storable, and intermittent due to changes in environmental conditions (Zhang *et al.*, 2017).

# 4    Relevance Cycle: Explorative interview study

The goal of this section is to examine the application domain of the CCR trading platform. Therefore, we conduct an explorative interview study with domain experts. In the following, we present our seven derived DR.

To ensure clients can obtain the overcapacity best matching their needs, they need to be able to switch between providers. To do so, *"we need standards so I can switch back and forth between different providers"* (Epsilon). Clients are usually bound to a single cloud infrastructure and must accept the providers' incompatible standards (Altmann *et al.*, 2008). Currently, this lock-in strategy affects small and medium-sized enterprises' competitiveness and market position since they deter from accessing the cloud market. Therefore, interoperability is needed, as *"in an interoperable Cloud environment customers will be able to compare and choose among Cloud offerings with different characteristics while they will switch between Cloud providers whenever needed without setting data and applications at risk"* (Loutas *et al.*, 2011), p.1). Thus, we propose

**DR1 (Interoperability):** *The platform should allow customers to switch services and providers conveniently.*

The platform must ensure a secure environment with guaranteed IT security and data security standards, including the security of the traded resources (Altmann *et al.*, 2008). The environment guarantees no viruses can spread, resource access is granted to permitted parties only, and all communication is encrypted (Veit and Gentzsch, 2008). As one interview participant mentioned, customers request to know the security standards of the services to understand *"under what security conditions can I use these resources"* (Beta). Thus, we propose

**DR2 (Security):** *The platform should provide a secure environment.*

Additionally, interviewees mentioned the need for transparency of the provided governance mechanisms. According to Bond (2015), untransparent governance mechanisms, such as price settings, causes negative feedback among stakeholders. Granting transparency to stakeholders goes along with less platform control by the intermediary, however more self-organizational effects and rapid growth in the user base (Hein *et al.*, 2016). In line with this information, the platform operator *"must provide transparency for the client, in all its governance mechanisms"* (Zeta) to attract and retain customers. Thus, we propose

**DR3 (Transparency):** *The platform should provide transparent processes.*

To attract participants, the platform operator *"has to be someone independent who does not try to put in any lock-ins through the back door, but it really has to be someone who is interested in the market and not in anything that he can then sell in this market"* (Delta). In line with this statement, the platform provider might be an independent third-party entity in the market, i.e., economically, and legally independent of other market participants. If the platform operator is part of one of the market sides, this might be viewed critically since it could result in higher market power and biased interests (Azer Bestavros, 2014). Thus, we propose

**DR4 (Interests):** *The platform should represent stakeholder interests equally.*

To ensure participants' trust in the platform, the provided services need to provide clear information about their quality (Cheng *et al.*, 2019). Restricted platform access and screening of services guarantee service quality. Furthermore, reputation mechanisms such as rating systems provide quality assurance services based on customer feedback. Such mechanisms install trust in the platform and the provided services (Cheng *et al.*, 2019). Hence, customers *"need a certain level of quality so they can trust the provided services"* (Lambda). Thus, we propose

**DR5 (Quality):** *The platform should ensure the quality of the services.*

Platforms are obliged to follow and continuously adapt to jurisdictional regulations of the respective countries they operate in, such as the general data protection regulations (GDPR) in the EU (Ponce, 2020). Regulations address issues such as data governance, fundamental rights, and accountability, just to mention a few (Belli and Zingales, 2017). While it is often challenging for platform operators to "navigate in the maze of obligations various jurisdictions place on them" (Belli and Zingales, 2017), p.127), compliance with jurisdictional regulations might also attract platform stakeholders since *"it is interesting for someone who lives in a non-GDPR country to make their data available under GDPR or use services under GDPR"* (Gamma). Thus, we propose

**DR6 (Jurisdiction):** *The platform must obey jurisdictional regulations.*

Additionally, interviewees mentioned service availability as a critical requirement for the platform. Customers need to be able to know *"that the services are reliable, that they are available when I want to use them"* (Alpha). Availability is essential *"so that the service users can count on them"* (Alpha). The importance of the availability of the resources is addressed by Altmann and Neumann (2009) and Buyya et al. (2009), who point out, that resource redundancy usually guarantees availability. Thus, we formulate

**DR7 (Availability):** *The platform should ensure the availability of overcapacity.*

To address our seven DR, in the following, we derive design principles.

# 5        Design Cycle: Deriving design knowledge

## 5.1       Awareness of the problem

Two geopolitical strategy decisions have mainly influenced the continuous existence of overcapacity: Europe's General Data Protection Regulation (GDPR) and the US's Cloud Act. GDPR defines strict legal terms regarding data privacy and data security in the EU, applicable to companies located in the EU as well as companies only operating in the EU (Wolford, 2021), such as traditional US cloud providers. However, since 2020 the US Cloud Act requires US companies to provide data stored in their data centers, whether abroad or in the US, if requested by the US government (Rojszczak, 2020), which contradicts the rules provided by GDPR. While many companies want to use the cloud to modernize their IT infrastructure, they fear for the security of their data when using data centers operated by traditional US cloud providers, such as Amazon Web Services, Microsoft, and Google (Kerkmann & Müller, 2020). Thus, GDPR, in combination with foreign geopolitical requirements, results in EU companies to prefer storing data locally in the EU independent of US hyperscalers, often using on-prem solutions (Kerkmann & Müller, 2020). Yet, they need to maintain their own on-prem infrastructure capacities, including overcapacity, due to a lack of EU-based hyperscalers. Recently, the EU initiative Gaia-X works towards those issues by enabling a European infrastructure to promote innovation through a secure data ecosystem based on European rules and regulations (Braud et al., 2021).

An MSP for CCR trading addresses the problem by enabling all companies to sell their overcapacity. In the relevance cycle, we conducted interviews with experts from industry and research. Based on the interviews, we identified seven DR: Interoperability, security, transparency, availability, interest, quality, and jurisdiction. Next, we derive DP from the energy market addressing the previously identified DR.

## 5.2       Suggestion

Reflecting on our results, we aim to suggest principles for designing an MSP for CCR trading based on existing theories in research from similar domains such as the energy market. We argue that the traditional energy market shares many characteristics with the cloud market. Traditionally, the energy market had few providers and many consumers, resulting in a central structure. With the emergence of decentralized energy resources, new active participants entered the electricity market. The occurrence of those distributed energy resources forced the traditional electricity market to adapt its market structure (Renewable Energy Agency, 2020). Different P2P electricity trading platforms have incorporated distributed energy resources and new market participants into the traditional electricity market (Renewable Energy Agency, 2020). P2P electricity trading is a business model where consumers, producers, and prosumers meet through an interconnected MSP, allowing platform participants to meet directly, without an intermediary (Park and Yong, 2017). Therefore, the electricity market serves as a field study for deriving DP. In the following, we present our six DP.

DR1 (Interoperability), DR3 (Transparency), DR4 (Interests), DR5 (Quality), and DR6 (Jurisdiction) stress the need for compatibility among CCR. Within the energy market, standards enable the interoperability of services, allowing customers to switch and combine providers easily. They also enable the ability to achieve cost-effectiveness, efficiency, and quality (Razazian and Yazdani, 2011). Standards define the basic requirements for (utility) services. In the energy market, standardization allows the global trading of distributed energy resources via platforms (Schweppe *et al.*, 1989). We propose to develop standards for CCR similar to the energy market to enable the interoperability of services. Therefore, we formulate the following initial design principle:

**DP1 (Compatibility):** *Provide the platform with technical, industry-specific, regulatory, and semantic standards in line with existing IT solutions in order to enable users to choose and switch between services and providers freely.*

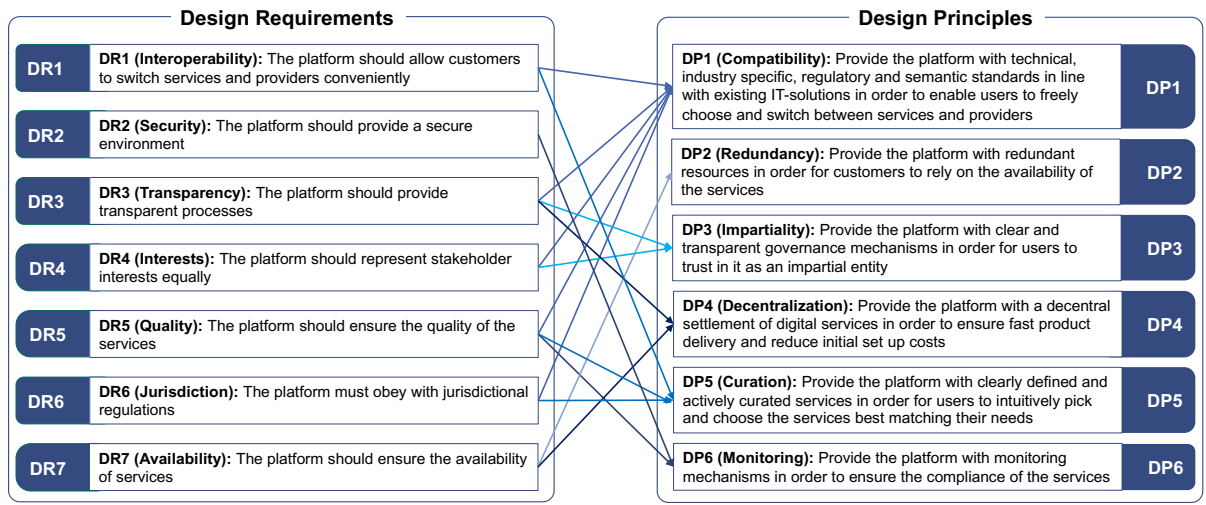

*Figure 3. Design principles and their relation to the respective design requirements*

DR7 (Availability) requests the reliable provision of CCR. However, the availability of CCR fluctuates due to changing peak load events. In return, this affects the service quality of the provided resources. Renewable energy resources are distributed, volatile, and intermittent in the energy market. They cannot adjust supply flexibly due to changing weather conditions. Energy trading platforms are connected to the primary power grid to balance sudden changes in supply or demand of distributed energy resources. The primary power grid takes on the role of a "residual balancer," guaranteeing a reliable energy supply (Ortega-Vazquez and Kirschen, 2010; Wilkins *et al.*, 2020). Like in the energy market, CCR is expected to be available dependably. To offer reliable CCR supply, we propose to connect the MSP for CCR trading with traditional cloud providers. Traditional cloud providers would take on the role of a "residual CCR balancer," providing redundancy to counterbalance fluctuations. Therefore, we formulate the following design principle:

**DP2 (Redundancy):** *Provide the platform with redundant resources in order for customers to rely on the availability of the services.*

DR3 (Transparency) and DR4 (Interest) are addressed by a platform representing the interests of its participants by making governance mechanisms comprehensible. In the energy field, digital platforms emerged, acting as exchanges for distributed energy resources (Kloppenburg & Boekelo, 2019). Energy trading platforms enabled new active consumers to enter the market. Those active participants expect transparency regarding energy characteristics and platform processes to actively participate in the energy market, such as negotiating prices based on energy types (Wilkins *et al.*, 2020). Furthermore, participants expect the platform to act in the best interest of the stakeholders rather than prioritizing commercial interests. Therefore, the platforms provide comprehensible and transparent processes, allowing participants to understand the represented interests of the platform (Smale and Kloppenburg, 2020). Exemplary platforms in the energy market are Piclo and Transactive Energy Initiative (Renewable Energy Agency, 2020). In line with the above findings, we propose establishing comprehensible platform mechanisms to ensure stakeholders actively participate and trust the platform. Therefore, we formulate the following design principle:

**DP3 (Impartiality):** *Provide the platform with clear and comprehensible governance mechanisms in order for users to trust in it as a non-partial entity.*

DR3 (Transparency) and DR7 (Availability) are addressed by a decentral setup of the cloud computing platform. Decentral platforms enable a direct service exchange between platform participants while simultaneously participants are affiliated with the platform. Within the energy market, decentralized digital energy trading platforms enable prosumers to meet while electricity trading occurs directly between parties. Such platforms allow prosumers to trade renewable energy at better prices, increase consumers' control of energy consumption, and enable easier renewable energy access (Renewable Energy Agency, 2020). We propose a decentral setup for the CCR trading platform where the platform only acts as an intermediator like in the energy market. Therefore, we formulate the following design principle:

**DP4 (Decentralization):** *Provide the platform with a decentral settlement of digital services in order to ensure fast product delivery and reduce initial setup costs.*

DR1 (Interoperability), DR5 (Quality), and DR6 (Jurisdiction) are addressed by clearly defined and curated services. In the energy market, energy platforms aim at providing diverse service offerings from different providers to customers. Despite the promising idea of establishing energy markets with retail choices, few people use the offer. Confusing service offerings and a lack of easy service comparison hinder customers from using the platform. Many different offers cause confusion and complexity to average customers (Zhai et al., 2019). In line with the above information, we suggest offering a clearly defined CCR to facilitate customers' purchasing process. Therefore, we formulate the following design principle:

**DP5 (Curation):** *Provide the platform with clearly defined and curated services in order for users to intuitively pick and choose the services best match their needs.*

DR2 (Security) and DR5 (Quality) are addressed by monitoring the provided cloud computing resources. In the energy market, certificates ensure the quality and compliance of renewable energy resources (Moser *et al.*, 2014). They specify, among others, what resources are utilized and how electricity service providers must comply (Lau and Aga, 2008). Based on the energy market's strategy to ensure compliance and quality of the services via certificates, we propose implementing certificates for idle CCR indicating the standards and actuality of the resources. Thus, we formulate the following design principle:

**DP6 (Monitoring):** *Provide the platform with monitoring mechanisms in order to ensure the compliance of the services.*

## 5.3 Development & Demonstration

We developed a generalized framework for CCR trading platforms based on insights from literature and our interview study, including the DP. We then instantiated the framework based on an analysis and discussion of the Gaia-X infrastructure and Gaia-X publications. In the following, we show how we develop and demonstrate the framework.

**Development:** The generalized platform framework is based on the decentral MSP structure by Hagiu and Wright (2015). Three types of entities are critical to the framework: consumers, producers, and the intermediating platform. We extend the framework by introducing "Standards". We hypothesize that standards turn CCR into a unit of trade by specifying, e.g., interoperability requirements. Furthermore, "Standards" verifies the identity of all network participants.

Stakeholders within the framework can interact in two ways: directly (consumer – producer) or indirectly (consumer – MSP – producer). Indirect interaction takes place to set up the initial contact between prosumers and negotiates the terms of the transaction. Direct interaction occurs once the deal is closed and the CCR is directly delivered from producer to consumer. Those two forms of interaction are critical features of any MSP (Hagiu and Wright, 2015).

The derived DP1 (Compatibility) and DP4 (Decentralization) affect the direct interaction between customers and producers. DP1 ensures the interoperability of traded services, allowing customers to integrate services from different providers. DP4 enables the direct exchange of services between customers and producers, independent of the platform. The latter allows faster product delivery and reduces complexity. Design principle DP6 (Monitoring) affects both the interaction of (consumer – MSP) and (producer – MSP).

The platform ensures service quality and compliance by probing the offered computing resources. DP2 (Redundancy) ensures sufficient availability of the services provided by the producers. Both relationships, (customer – MSP) and (producer – MSP), are based on a contractual agreement. The relationship between (MSP – standards) and (prosumer – standards) are based on a verification contract. A verification contract refers to a contractual agreement without commercial interests.

DP3 (Impartiality) and DP5 (Curation) address the features of the platform. The platform provides a neutral environment representing the interests of the stakeholders. Further, curation of the offered services facilitates service selection and the finalization of the transaction.

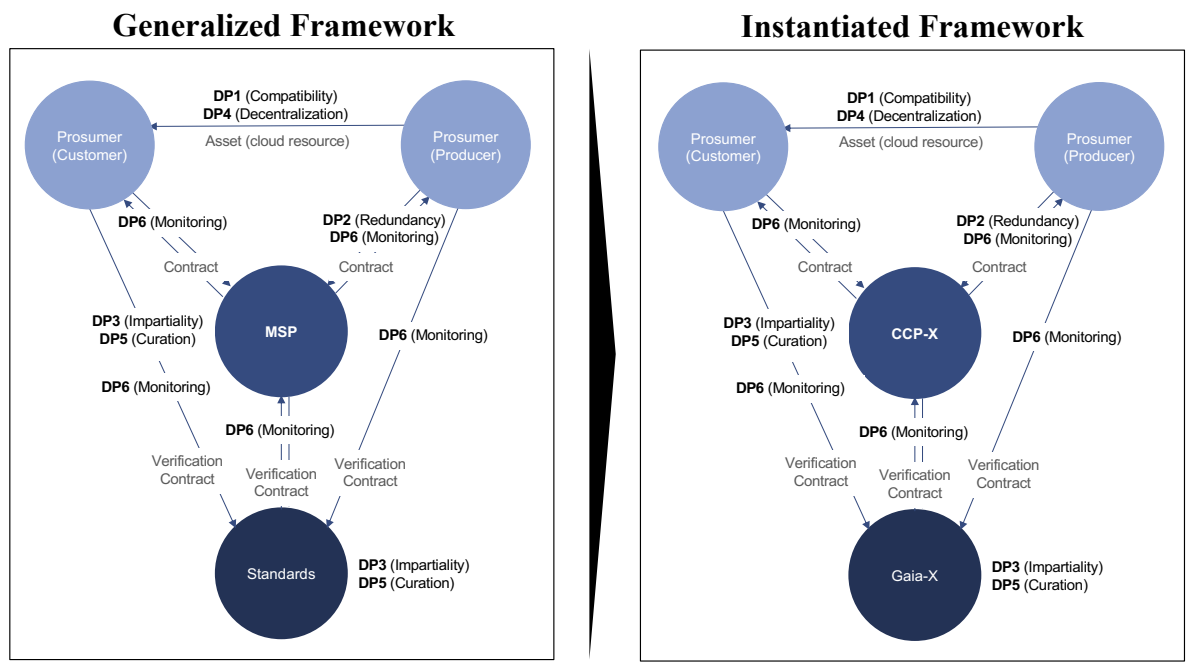

*Figure 4.*     *Generalized platform framework (left); Instantiated framework—"Cloud Computing Platform enabled by Gaia-X (CCP-X)" (right)*

**Demonstration:** The European cloud market actively strengthens its international competitiveness by developing a standardized regulatory framework called Gaia-X. Gaia-X works towards open, independent, transparent, and interoperable technology standards (Braud et al., 2021). The aim is to enable an infrastructure and data ecosystem based on European values and regulations. The envisaged ecosystem should allow cloud services users' digital sovereignty and European cloud providers to scale and compete. Gaia-X, at its core, is based on four federated services: federated catalog, identity and trust, data sovereignty services, and compliance. The federated catalog offers an index repository of Gaia-X self-descriptions to enable the discovery and selection of providers and service offerings. The self-descriptions compile information about the provider and the service offering. Identity and trust provide authentication and authorization, credentials management, decentralized identity management, and the verification of analog credentials. Data sovereignty services provide data exchange and usage agreements. Compliance offers mechanisms to ensure participants' adherence to rules and regulations (Braud et al., 2021).

We take Gaia-X as the foundation for hypothesizing the platform CCP-X based on our generalized framework. CCP-X is an MSP certified by Gaia-X, offering curated CCR overcapacity. Customers interested in and providers of idle CCR meet on the platform to exchange service information, agree on pricing, and sign contractual agreements. Once an agreement is signed, the exchange of CCR occurs between customer and producer directly, as mentioned by Hagiu and Wright (2015). Any company can take on the role of consumer or providers. Hence, CCP-X fosters prosumption. CCP-X and CCP-X's stakeholders have a direct relationship with Gaia-X. Gaia-X authenticates and identifies the platform and its stakeholders, making them "trustworthy" business partners. After being certified by Gaia-X, the stakeholders can then trade on CCP-X. Thus, all services offered by CCP-X and all stakeholders interacting with the platform are certified by Gaia-X. CCP-X depicts a use case that can be implemented on Gaia-X's federation services within a Gaia-X consortium. Figure 4 depicts the generalized platform framework (left) and the CCP-X framework (right), including the respective DP.

## 5.4 Evaluation

In our summative ex-post evaluation, we assessed each DP in detail. The experts perceived all proposed DP as valuable and effective in solving the identified DR. Participants especially highlighted the demand by customers for compatibility of services provided by different providers (DP1 Compatibility). This is in line with a study by Gartner stating that 81% of public cloud companies already follow a multi-cloud strategy, meaning they obtain services from different providers (Goasduff, 2019). As of now, they do so mainly to avoid vendor lock-in and to decrease security risk. Additionally, dispersing tasks running in the cloud over multiple cloud providers instead of a few hyperscalers allows minimizing the risks of a single provider failure or a single vendor lock-in. Compatibility standards would allow customers to switch freely between providers independent of any vendor lock-ins (Różańska and Kritikos, 2019).

Domain experts suggested two additional aspects. First, we discussed the applicability of CCP-X to other services such as trading in bundles of CPU and artificial intelligence. Trading in bundles would allow providers to offer infrastructure services linked to applications and software services. Second, we discussed the implementation of CCP-X within a Gaia-X consortium. The domain experts thought it feasible to implement CCP-X as a use case within a Gaia-X consortium. The results of our evaluation confirm the usefulness, the effectiveness, and the potential for implementation of our derived design knowledge and platform framework.

## 6 Discussion and Conclusion

In the following, we discuss the implications and limitations of our work.

**Managerial Implications.** From a platform perspective, the successful implementation of MSP depends, among other factors, on the number of active users on the platform generating network effects. There are two types of network effects, direct network effects, and indirect network effects. Direct network effects increase the number of users on the same side (demand or supply side), while indirect network effects cause the number of participants in the other user group to rise (Evans and Schmalensee, 2019; Hagiu and Wright, 2015). Initially, MSPs are faced with the challenge of the so-called chicken and egg dilemma, meaning consumers hesitate to interact with the platform unless enough producers offer their services and vice versa. Since our platform framework enables participants to take on the role of consumers as well as producers, this initial hurdle could be deemed less severe since prosumers have a greater impact on network effects than consumers or producers, as they participate on both sides of the platform. Therefore, prosumption is critical to the platform since it might positively impact initial engagement on the platform. Hence, it would increase the chance of successful implementation of the proposed platform framework.

**Political Implications**. The proposed platform framework opens the cloud market to new entrants distinct from existing players by providing a platform approach that allows any company to sell CCR overcapacity. This platform can have major policy implications on general CCR adoption and independence. From a geopolitical perspective, the platform could strengthen national cloud ambitions that are independent of mostly American hyperscalers and thereby create a balance. For example, from a European perspective, the instantiated version of the proposed platform could strengthen the position of European cloud providers by providing a central platform for offering CCR capacity in line with European regulations and enabled through Gaia-X. A central platform would also provide easy access for European companies to acquire their resources. All platform participants including the platform itself would be certified by Gaia-X. To date, CCR are provided by different dispersed vendors, making it difficult for customers to evaluate and find the right CCR (Loutas *et al.*, 2011). Providing CCR complying with European regulations is in line with Europe's plan for digital sovereignty. Currently, this plan is based on two pillars: cloud sovereignty and data sovereignty. Cloud sovereignty refers to the need for cloud services that comply with European regulations. Data sovereignty will enable participants to share data safely in a consortium. The European initiative Gaia-X works on the implementation of Europe's digital sovereignty plan (Braud et al., 2021). Hence, our trading platform for CCR overcapacity could support the political interest of Europe for digital sovereignty.

**Theoretical Implications.** To date, as our structured literature review shows, research on multi-sided platforms for CCR offers few recent and evaluated insights into design knowledge for CCR trading platforms. We address this research gap by conceptualizing a platform framework for CCR trading based on seven DR derived from practice and six DP derived from the energy market. Deriving DR from practice allows us to base our research on current market needs.

First, our derived design knowledge extends the existing body of knowledge by providing the first recent qualitative evaluation of DP and a platform framework for CCR trading platforms. We evaluated our DP and their relation to the respective DR with domain experts from the Gaia-X steering committee. We also evaluated the feasibility of the platform framework with domain experts. The domain experts thought it is feasible to implement the platform framework based on Gaia-X. Hence, we contribute to the existing body of knowledge by conceptualizing a general platform framework for CCR trading evaluated by domain experts which can also be instantiated as a Gaia-X use case. Second, we provide design knowledge, which might be applicable to other markets trading resources with similar characteristics to CCR. For example, we may have gained insights that can be useful for energy markets as well. DP5 (Curation) might help to attract more prosumers to energy trading platforms. To date, only a few people use energy trading platforms due to confusing service offerings and the lack of easy comparison of services (Zhai et al., 2019). Curation might help decrease confusion and complexity for the average customer. The design knowledge might further apply to other energy markets. In summary, we present recent and evaluated design knowledge as well as a platform framework for a CCR trading platform. This platform might also be applicable to other markets. Hence, our research contributes new and evaluated insights to the design knowledge base of CCR trading platforms.

**Limitations and Outlook.** Although we conducted our DSR project according to accepted guidelines, there are some potential limitations to our research. We evaluated the design knowledge and the resulting framework with a limited number of domain experts. Additional evaluation episodes would strengthen the significance of our results. Regarding the development of design features, we see potential in a more detailed elaboration of the design principles by applying it to specific disciplines like mechanism design for the appropriate allocation mechanisms or computer science for the delivery protocols and the development of the infrastructure. In addition, the development of new classes of services, such as the trading of bundles of infrastructure services and related applications, e.g., AI, leaves room for further research. Investigating the implementation of the conceptualized platform framework on the Gaia-X infrastructure should expand our research. Further research should develop a business model describing how the platform would capture value, e.g., define revenue streams, develop relationships with prosumers, and outline pricing structures. Thereby, application criteria critical for Gaia-X should be considered. Lastly, in terms of generalization, we see potential in the applicability of the proposed platform framework to other markets trading resources with characteristics similar to CCR.